\UseRawInputEncoding
\documentclass[prl,aps,twocolumn,groupedaddress,floats,nopacs]{revtex4}
\usepackage{graphicx}
\usepackage{dcolumn}
\usepackage{bm}
\usepackage{color}
\def\he4{$^4$He}
\def\hel3{$^3$He}
\def\Am3{\AA$^{-3}$}
\def\beq{\begin{equation}}
\def\eeq{\end{equation}}

\begin{document}

\author{Anatoly Kuklov$^1$, Emil Polturak$^2$, Nikolay Prokof'ev$^3$ and Boris Svistunov$^{3,4,5}$}
\affiliation{$^1$ Department of Physics and Astronomy, CSI, and the Graduate Center of CUNY, New York}
\affiliation{$^2$ Faculty of Physics, Technion-Israel Institute of Technology, Haifa, Israel}
\affiliation{$^3$ Department of Physics, University of Massachusetts, Amherst, MA 01003, USA}
\affiliation{$^4$ National Research Center ``Kurchatov Institute," 123182 Moscow, Russia}
\affiliation{$^5$ Wilczek Quantum Center, School of Physics and Astronomy and T. D. Lee Institute,
Shanghai Jiao Tong University, Shanghai 200240, China}

\title{Anomalously Small Excitation Gaps as a Precursor of Dislocation Core Superfluidity in Solid Helium-4}

\date{\today}
\begin{abstract}
In the vicinity of the insulator-to-superfluid quantum phase transition in its core, a dislocation
in a He-4 crystal supports particle-hole excitations with arbitrary small gaps. These exotic analogs
of Frenkel interstitial-vacancy pairs should manifest themselves in various threshold and thermoactivation
effects. In Worm Algorithm simulations, we reveal the presence of corresponding small
gaps via anomalous thermoactivation behavior of particle number fluctuations, which we unambiguously associate
with dislocations by ``visualization'' techniques. Experimentally, the related threshold
and thermoactivation dependencies could be observed in the ultrasound absorption.
\end{abstract}

\maketitle

While being one of the most studied strongly interacting many-body systems, solid \he4 remains enigmatic due to the unique and extremely
unusual behavior of its topological defects.
An early fundamental interest to solid \he4 has been sparked by the proposal that it is a supersolid \cite{Gross,Andreev,Thouless,Chester}---a state of matter combining properties of solid and superfluid. However, subsequent large-scale experimental efforts yielded no supportive evidence. Huge interest to the subject has reemerged after the experiments have found a possible evidence of the supersolid phase in the temperature variation of the torsional oscillator frequency  \cite{KC}. This variation, however, was later shown to be unrelated to the supersolidity and was explained in terms of the dislocation dynamics \cite{Beamish_Nature}.

In the absence of reliable analytical tools for studying solid \he4, {\it ab initio} numerical methods based on the path integral formulation of quantum mechanics  (see in Ref.~\cite{Cep1995}) became especially important. Immediately after the observation of the torsional oscillator anomaly, several groups have unambiguously established that perfect \he4 crystal is not a supersolid \cite{Cep,WA,Clark}.

Almost simultaneously with ruling out the supersolid phase of perfect \he4 crystal, {\it ab initio} simulations revealed superfluid properties
of the structural defects. Certain grain boundaries \cite{GB} and screw dislocation (with Burgers vector along the hcp symmetry (C-) axis) \cite{screw}  have been found to support (quasi) 2D and 1D superfluidity, respectively.
About two years later, a flow of \he4 atoms through solid \he4 has been detected in the experiment conducted by the UMass group \cite{Hallock}. Typical characteristics of this flow---temperature and bias dependencies---are clearly inconsistent with any type of classical dissipative dynamics. The flow rate shows subohmic dependence on the bias and it decreases as temperature increases. These features have been confirmed by other groups \cite{Beamish,Moses}.
Furthermore, it was discovered that the flow observed in Refs.~\cite{Hallock, Beamish} is accompanied by yet another
unusual (an absolutely unexpected) feature---the so-called {\it syringe effect}, when matter accumulates inside the solid biased by the chemical potential. Later, the syringe effect has been attributed to the phenomenon of {\it superclimb} of edge dislocations (with their Burgers vector oriented along the C-axis) observed in the {\it ab initio} simulations \cite{sclimb}.

Current understanding of the supertransport through solid (STS)  \cite{Hallock,Moses} and the syringe \cite{Hallock,Beamish} effects is based on the assumption that the solid \he4 contains a network of dislocations with superfluid cores \cite{shevchenko} providing pathways for the superflow. There is, however, no known direct way to image such a network in solid \he4 (in contrast to other materials). Here we suggest an alternative approach to the problem.

The STS effect vanishes at pressures only few bars above the melting line.
Within the picture of the superfluid dislocations network,  it is natural to assume that vanishing of the STS effect is caused by the
superfluid-to-insulator quantum phase transition in the dislocation core. The insulating state is characterized by gapped particle-hole excitations. In 1D, the gap, $\Delta$, develops continuously and, therefore, close to the transition it should be anomalously small if compared with typical excitation energies for a vacancy, $E_v=(13.0\pm 0.5)$K, and interstitial, $E_i=(22.8\pm 0.7)$K,  in the perfect  \he4 crystal \cite{fate}. This gap $\Delta$ can be seen directly in the thermally activated response characterized by some finite activation energy $E_a \sim \Delta$, and, most importantly, it should emerge from zero at the densities, where the STS effect is observed, to a finite value at densities where the STS effect seizes to exist.

Here we present the results of first-principles simulations of several edge dislocations characterized by different orientations of their Burgers vectors and cores, with the goal to reveal small $E_a$'s and demonstrate that these are due to particle-hole excitations living on dislocations. Among our samples there is one dislocation that has a superfluid core close to the melting density (0.0287\AA$^{-3}$). It is the partial dislocation, which is a boundary of the E-fault in the hcp crystal \cite{Hirth}. At the density 0.0300\AA$^{-3}$ it is insulating with $E_a \approx 0.7$K. This value is significantly below the values obtained for the dislocations that demonstrate no superfluidity at low densities.

{\it Theoretical framework.} Within the grand canonical ensemble (GCE) at a fixed value of the chemical potential $\mu$, the expectation value of the total number of particles $\langle N \rangle $  in a perfect crystal changes as a function of temperature according to the relation  $\langle N \rangle = N_0 - N_v + N_i $, with $N_{v,i} \sim N_0 \exp(-E_{v,i}/T)$ being the mean  numbers of vacancies and interstitials, respectively, and $N_0$ stands for $\langle N\rangle $ at $T=0$ 
(we neglect the pre-exponential temperature dependence arising at temperatures well below the tunneling dispersion bandwidth).  
If a system is characterized by $E_v \neq E_i$ due to lack of the particle-hole symmetry it is possible to measure the point defect number characterized by the smallest activation energy  in the main exponential approximation, provided $E_i$ and $E_v$ are not too close to each other. This is the case in a perfect \he4 crystal \cite{fate}. The relation between $E_i$ and $E_v$ close to a dislocation is not known {\it a priori}. Furthermore, the very notions of vacancies and interstitials become poorly defined because when particles are added (or subtracted) to (from) the dislocation core the result is the dislocation climb \cite{Hirth}. In fact, adding or removing a particle to the core of edge dislocation is equivalent to creating a jog-antijog pair at its minimal possible separation. It is clear then that activation energies
for adding and removing particles are the same---by the token of the symmetry with respect to the direction of the climb. This implies the emergence of the particle-hole (interstitial-vacancy) symmetry in the core of the dislocation resulting in the cancellation of $N_i$ and $N_v$ in $\langle N \rangle$.

Possible cancellation of $N_v$ and $N_i$ forces us to resort to the mean squared fluctuations of $N$ instead:
\[
\sigma^2_N = \langle N^2 \rangle  - \langle N\rangle^2 .
\]
For this observable, the contributions of both types of defects come with the same sign.
At low density (and temperatures much higher than a possible degeneracy temperature) point defects 
behave as an ideal gas and one can rely on the simple relation
\beq
\sigma^2_N=N_v + N_i.
\label{NvNi}
\eeq

{\it Samples and simulations.}
Initial samples  have been prepared starting from atomic positions arranged according to the perfect hcp symmetry. In order to produce a topological defect, a corresponding half-plane of atoms has been removed and the remaining spatial gap healed by means of purely classical simulations with some repulsive interaction potential between atoms. Then, simulations have been conducted by the Worm Algorithm \cite{WA} for several temperatures starting from T=0.25 K and up to 2.5K at corresponding values of the chemical potential $\mu$ and two different  densities $0.0288$\AA$^{-3}$ and $0.030$\AA$^{-3}$. 
Quantum configurations of the worldlines were periodically projected into time-averaged 
classical positions. As a qualitative assessment of the role of quantum fluctuations, the map of particle exchanges has been superimposed atop of the classical snapshot of the positions. This map has served as an imaging tool for the most probable areas where particles were introduced or removed. The details of the imaging protocol were described in detail in Ref.~\cite{us}.
Several types of dislocations have been simulated: (i) basal edge dislocations (with both Burgers vector and the core belonging to the basal plane), (ii) non-basal edge dislocation topologically equivalent to a jog of the basal dislocation, and  (iii) partial dislocation corresponding to the boundary of the E-fault \cite{Hirth}.

{\it Basal fault edge dislocation.}
\begin{figure}[!htb]
\vskip-8mm
	\includegraphics[width=1.1 \columnwidth]{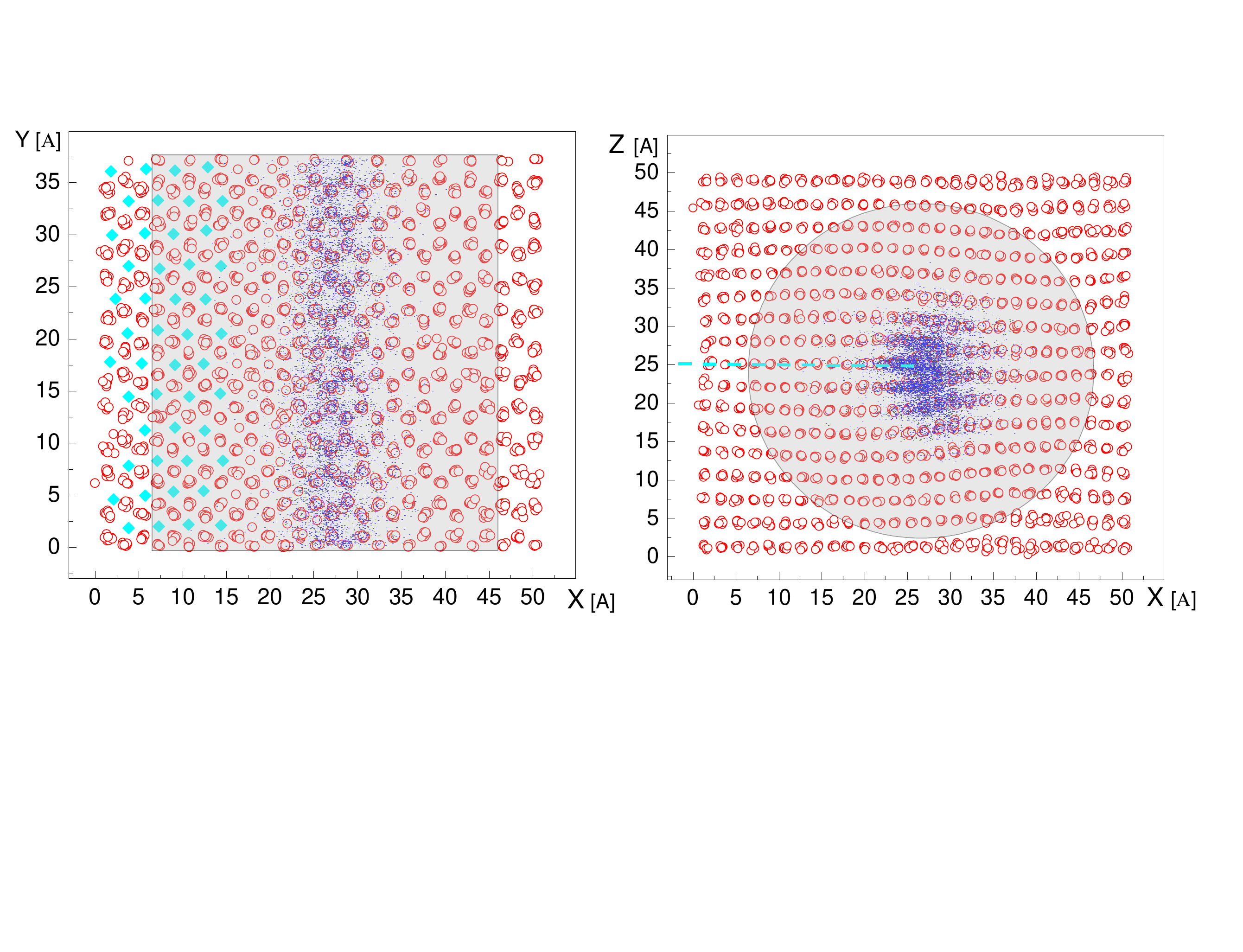}
	\vskip-12mm
	\caption{(Color online) Atomic positions (open circles) of a sample containing the E-fault edge dislocation. The exchange map (blue dots) visualizes the space-time regions contributing to the particle number fluctuations. The dimmed areas indicate the cylinder containing fully updatable particles. Particles outside this cylinder are frozen. Left panel: Columnar view along the hcp axis. The incomplete fault plane is partially shown by (cyan) diamonds. Right panel: Columnar view along the dislocation core (along the Y-axis). The incomplete fault plane is marked by the (cyan) dashed line. }
	\label{fig11a}
\end{figure}
The E-fault in the hcp structure is formed when in the  ABABABAB... stack of triangular atomic layers forming the hcp structure 
one inserts a different triangular layer C, say, ABABCABAB... \cite{Hirth}. The ABC element of 3 layers has the fcc symmetry \cite{Hirth}.
 If the plane C is incomplete, the edge of the C-plane represents one partial of the full edge dislocation with the Burgers vector along the hcp symmetry axis. This full dislocation has a superfluid core at low temperature---as observed in the simulations \cite{sclimb}. Furthermore, it splits into two partials one of which is the edge of the E-type fault. Here we study the thermal properties of this partial at density $0.03 \AA^{-3}$---when there is no significant superfluid response along the core. (We do not discuss the other partial dislocations---edges of different basal faults; see in Ref.~\cite{Hirth}.) A typical snapshot of the atomic positions of the E-fault partial dislocation is shown in Fig.~\ref{fig11a}.

We asses the superfluid response via the Luttinger parameter $K$, which we extract from the fluctuations of the
worldline winding number along the dislocation core, $W_x$:
\[
K\, =\, \pi \sqrt{\sigma_N^2 \, \langle W_x^2\rangle} .
\]
The procedure is identical to that used in Ref.~\cite{screw}. The values $K>2$ correspond to the superfluid
ground state. In a weakly insulating sample of essentially finite linear size $L$, the value of $K(L)$ is somewhat smaller than $2$, which, in accordance with the theory of superfluid-to-Mott-insulator transition in 1D, implies that $K(L)\to 0$ at $L\to \infty$.


In order to prevent the annihilation of dislocations with the opposite ``charges,'' full quantum updates have been applied to only one dislocation---inside the cylinder shown in Fig.~\ref{fig11a}. This was achieved by freezing out atoms (together with the ``antidislocation'') outside the cylindrical region. These  particles have been annealed first as if they were distinguishable quantum particles
 (by excluding exchange cycles) and, then, have their worldlines frozen for the rest of the simulation.  As a result, the periodic boundary conditions were satisfied only along the dislocation core inside the cylinder (the Y-axis in Fig.~\ref{fig11a}).

\begin{figure}[!htb]
	\includegraphics [trim = 70 0 75 50, width=1 \columnwidth]{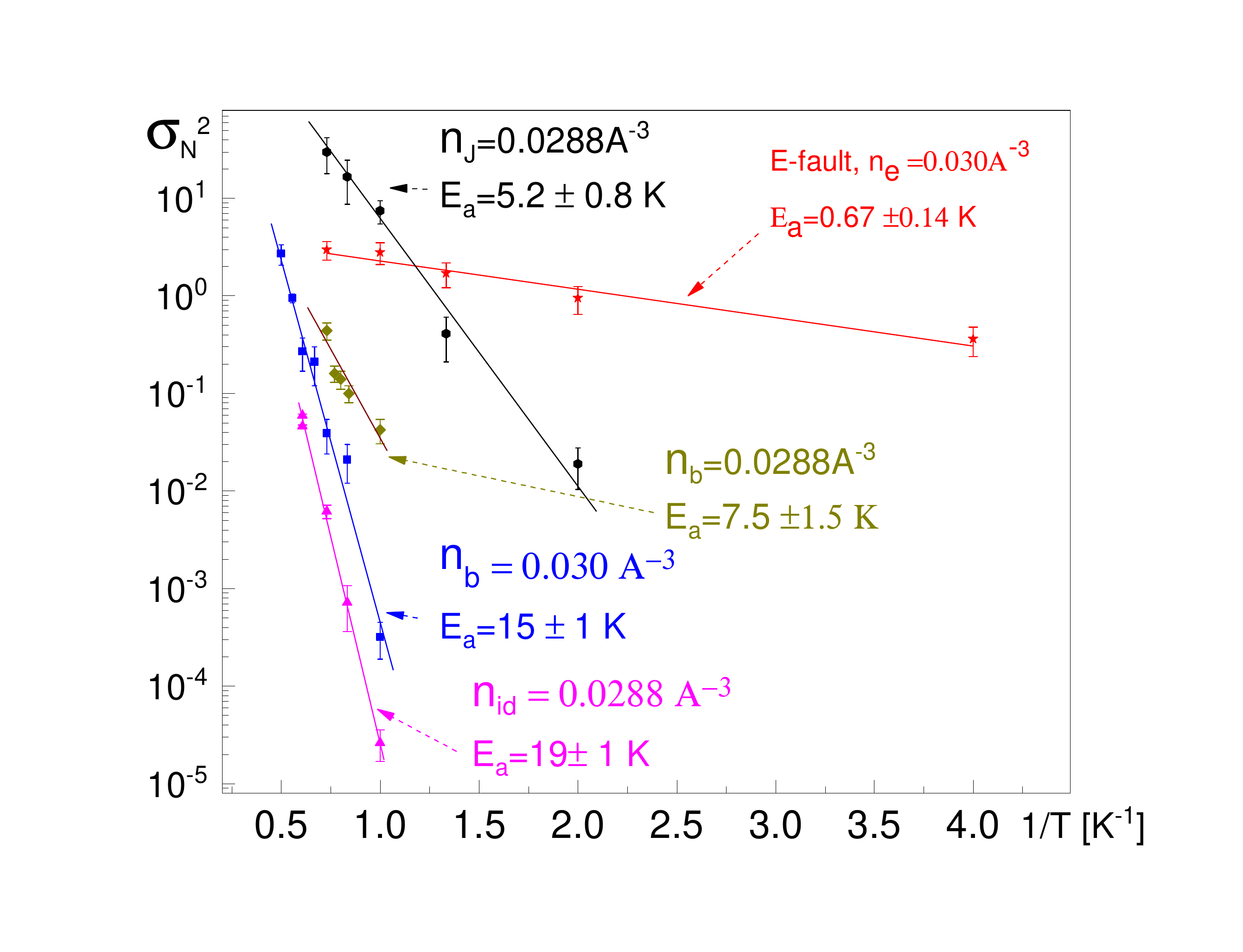} 
	\caption{(Color online) Particle number fluctuations $\sigma_N^2$ determined for: a pair of basal dislocations at two densities $n_b =0.0288 \AA^{-3}$ (dark yellow rhombi), $ 0.030 \AA^{-3}$ (blue squares); perfect hcp crystal of density $n_{id} = 0.0288 \AA^{-3}$ (pink triangles); the E-fault partial at density $n_e =0.030 \AA$ (red stars); a pair of jogs (black hexagons) at density $n_J=0.0288 \AA^{-3}$. The corresponding activation energies and the densities are mentioned next to each line.}
	\label{fig10}
\end{figure}

{\it Thermoactivation responses of the dislocations.}
All samples---including the one without dislocations---have demonstrated the thermoactivation behavior. The results for $E_a$ are summarized in Fig.~\ref{fig10}. The smallest value for non-superfluid dislocations (i) and (ii) is found to be $E_a=5.2 \pm 0.8$K for the jog-type dislocation (ii). This value agrees well with the purely classical estimate of the jog energy 5.8K \cite{Tsuruoka}.
The value of $E_a\approx 0.7$K for the E-fault dislocation---seen at appropriately low temperature---is very small compared to the values for other dislocations. The Luttinger parameter for this sample is slightly smaller than $2$, meaning that the core of the dislocation is in a weak insulating state, which explains the anomaly.

{\it Discussion.}  
We interpret  all the found responses from the perspective of the excitation gap suppression by local elastic strain introduced in Ref.\cite{strain}.
According to it, local deformations around dislocations or any other defect suppress the energy required to create a free vacancy.
Once the local strain reaches 10-12\% close to the melting density, such energy becomes zero, and this explains the formation of the superfluid core for the dislocations with Burgers vector along the C-axis \cite{screw,sclimb}. In contrast, other dislocations are characterized by strains insufficient for closing the energy gap, which explains their insulating character (with $E_a \geq 5-6$K) even close to the melting. 

Our main result is revealing an anomalously low activation energy, $E_a \approx 0.7$K, for a dislocation featuring superfluid core at melting density that turns insulating at higher density. Accordingly, $E_a$ changes from $E_a=0$ in the superfluid phase to a finite value $E_a$ in the insulating state.

Experimentally observing $E_a \sim 1$K with such a characteristic density dependence  would provide a crucial evidence supporting the explanation of the STS effect in terms of superfluidity of dislocation cores.
We suggest that  ultrasound studies of ultra pure solid \he4 might prove instrumental in this respect. There are two mechanisms which can contribute to the sound absorption. One corresponds to the sound attenuation by a cloud of normal excitations around the dislocation core. A detailed mechanism of how low-gap excitations
around dislocation cores can modify the attenuation of sound
has been described in Ref.~\cite{us}.  In this case, the low-$E_a$ dissipation mechanism 
will show up as a peak in the temperature dependence of the absorption coefficient  
at $T\approx E_a$, with the logarithmic corrections dependent on the sound frequency.

One can also think of another---conceptually, much more straightforward, but apparently more challenging from the technical point of view---experimental method based on the absorption of higher frequency ultrasound at temperatures
$T \ll \Delta$. When the ultrasound frequency $f$ exceeds the gap $\Delta$, the excitations can be created directly by the sound wave
as opposed to scattering of existing ones (cf. Ref.~\cite{SC}). By the Fermi Golden Rule, the absorption coefficient $\alpha(f)$ should have the following characteristic profile featuring a sharp peak at $f = \Delta$:
\beq
\alpha(f) \propto \left\{ {\begin{array}{*{20}{c}}
(f - \Delta )^{ - 1/2}, \quad \mbox{if} \quad f > \Delta \\
\qquad \qquad 0, \qquad \mbox{if} \quad f < \Delta
\end{array}} \right. \qquad  (T=0) .
\label{threshold}
\eeq
This type of measurements requires  $f \approx 10\div 20$GHz for $\Delta \sim 0.5\div 1$K.  Since the gap develops continuously, it can, in principle, be detected at lower frequencies closer to the transition threshold. However, the temperature must be lowered accordingly, to guarantee the $T\ll \Delta$ condition.

The method of Ref.~\cite{SC} is based on creating a hypersound in a mechanical resonator illuminated by high intensity light. The produced sound must then be transferred to the solid \he4. Since, the attenuation length of the hypersound is small, this setup may be quite challenging to realize. [In particular, some heating of the resonator can be problematic to mitigate]. Thus, we suggest  an alternative approach to achieving the same goal---producing hypersound in the required range of frequencies right inside a sample of solid \he4.  It is based on injecting ions into the solid \he4 and applying a microwave electromagnetic radiation (in the range $f \approx 10\div 20$GHz). Ions tend to condense on dislocations and grain boundaries in solid \he4,  which results in the ionic-current bursts observed in Ref. \cite{ions}: the mobile dislocation lines carried ions and, accordingly, were driven by an external electric field. The cell containing solid \he4 would be made as a part of a microwave resonator so that the injected ions can be probed using microwaves. In their turn, the ions bound to dislocations will excite mechanical  waves along the dislocations. The damping of the electro-mechanical waves should increase once their frequency exceeds the smallest between the intrinsic gap $\Delta$ and the binding energy $E_{\rm ion}$ of the ions to dislocations. Since $\Delta \sim 1$K and $E_{\rm ion}$ exceeds 10-20 K \cite{ions}, the shape of the absorption line is expected to be described by Eq.(\ref{threshold}) at frequencies well below 200 GHz. One important advantage of this method is that using wideband microwave sources should make it possible to look for the gapped dislocations over a wide range of values of $\Delta$.

At this juncture it is instructive to briefly review the present status of the studies of the thermal activation in solid \he4. These studies have a long history of reporting various activation energies. In solid \he4 contaminated by $^3$He the processes of the impurities binding (and unbinding) to dislocations are characterized by $E_a$ about $0.7\div 0.8$K  \cite{Polturak_3_6} (see also Ref.~\cite{RMP}). More recent studies using ultrasound have reported smaller values---ranging from 0.18K to 0.35K  \cite{Iwasa} for the $^3$He binding energy.
There are also reports of the activation behavior of the intrinsic origin -- unrelated to $^3$He impurities. In Ref.~\cite{Tsuruoka}, the activation energy about $12 \div 14$K has been reported. In Refs.~\cite{Beamish_5,Polturak_3_6}, the values $E_a \sim 5 \div 6$K and 3K \cite{Polturak_3_6} were found. 

Ultrasound studies  of ultra pure solid \he4 (with 1.5ppb of $^3$He) \cite{Goodkind} have  revealed the absorption peak characterized by the activation energy 0.7K at crystal densities corresponding to melting. This value increases linearly with density and reaches 1.18K at the density only a small fraction of a percent above the melting.
It was interpreted in terms of the energy gap in the system of Bose-condensed excitations, and considered as evidence for a possible  supersolidity of a perfect \he4 crystal \cite{Goodkind2002}. 
While this interpretation is ruled out by recent experiments and simulations, it remains to be seen  
to what extent this feature might be attributed to the anomalously low gaps for weakly insulating dislocations.
More recently the ``anomalous'' absorption peak at $T=0.7$K peak has also been found in Ref.\cite{Iwasa}. The origin of such a peak is unknown.
The ultimate understanding of the origin of anomalously small gaps and absorption peaks may be achieved with an 
experimental setup allowing to study both the excitation gaps and the DC supertransport within one and the same sample.


{\it Acknowledgments.}
We thank I. Iwasa for useful discussions of the anomalous absorption peak \cite{Iwasa}.
 This work was supported by the National Science Foundation under the grants DMR-1720251, DMR-2032136 and DMR-2032077.  Computations have been conducted at the ATLAS Israel Group supercomputing cluster. We acknowledge assistance by its system administrator David Cohen.

\end{document}